# Beam-spread determination for luminosity measurement at CEPC

I.Smiljanic[a,1], I. Bozovic Jelisavcic[a,*], G. Kacarevic[a]

[a]Vinca Institute of Nuclear Sciences, University of Belgrade 11000 Belgrade, Serbia

**Abstract:** Any asymmetry in energy of the colliding beams will lead to a longitudinal boost of the center-of-mass frame of colliding particles w.r.t. the laboratory frame and consequently to the counting loss in luminometer due to the loss of colinearity of Bhabha final states. At CEPC running at the $Z^0$ pole, asymmetry in energy of the colliding beams should be known as well as 12.5% of the beam-spread, in order to control the uncertainty of Bhabha count at the level of $10^{-4}$. Here we discuss the method, initially proposed for FCCee, to determine variation of the beam-spread from the measurement of the effective center-of-mass energy in $e^+e^- \to \mu^+\mu^-$ collisions.

## 1. Introduction

The Circular Electron Positron Collider (CEPC) is a large international scientific facility proposed by the Chinese particle physics community in 2012 for precise measurements of the Higgs boson properties, as well as for precision EW physics. These measurements should provide critical tests of the underlying fundamental physics principles of the Standard Model (SM) and are vital in exploration of new physics beyond the SM. In electron-positron collisions, the CEPC is designed to operate at around 91.2 GeV as a *Z* factory, at around 160 GeV of the *WW* production threshold, and at 240 GeV as a Higgs factory. The vast amount of bottom quarks, charm quarks and *τ*-leptons produced in $Z^0$ decays also makes the CEPC an effective *B*-factory and *τ*-charm factory [1].

In order to achieve precision required for realization of the CEPC physics program, relative uncertainty of the integrated luminosity measurement should be of order of $10^{-4}$ at 91.2 GeV and of order of $10^{-3}$ at 240 GeV. The method for integrated luminosity measurement at CEPC is described in [1 and 2]. As discussed in Sec.2, asymmetry in energy of the colliding beams should be known as well as 12.5% of the beam-spread, in order to control the uncertainty of Bhabha count at the level of $10^{-4}$. In this paper we present a method to experimentally determine the beam-energy spread, from the measured effective center-of-mass energy *s'* in $e^+e^- \to \mu^+\mu^-$ collisions. The method is initially proposed for FCCee [3]. We quantify sensitivity of *s'* to the beam-spread at 91.2 GeV and 240 GeV CEPC, in the presence of additional effects like ISR and finite detector resolution. We also determine a time required to achieve experimental control of the beam-spread corresponding to the $10^{-4}$ relative uncertainty of the integrated luminosity (Sec. 4).

## 2. Requirements on beam energy spread

According to [1], beam energy spread at CEPC will not exceed 0.134% of the beam energy at 240 GeV center-of-mass and 0.08% of the beam energy at 91.2 GeV, and its shape will ideally be Gaussian. That implies that the difference in energy of colliding particles can be as large 322 MeV for the Higgs factory and up to 73 MeV for the $Z^0$ factory, which gives a rise to a longitudinal boost of the center-of-mass (CM) frame of colliding particles with respect to the lab frame, $\beta_Z$:

$$|E_+ - E_-| = \Delta E \quad \to \quad \beta_Z = \frac{\Delta E}{E_{CM}}$$

---

[1]E-mail: i.smiljanic@vinca.rs
* speaker

The above further leads to counting loss in luminometer, due to acolinearity of Bhabha final states, as shown at Figure 1. Uncertainty of count of $10^{-4}$ implies knowledge of the asymmetry in beam energies at the level of 12.5% of the beam-spread at the $Z^0$ pole. As this requirement is below the natural energy spread of the beam, a dedicated method has been applied in order to determine the effective beam-spread from the measured effective center-of-mass energy, in a relatively short time interval.

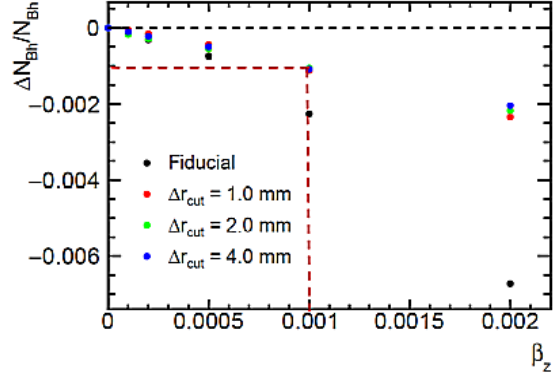

**Figure 1.** Counting loss in the luminometer due to longitudinal boost of the CM frame. Δr values correspond to different polar angle acceptance of detector left and right arms. Dashed line indicates $10^{-3}$ uncertainty in count.

It is interesting to note that other EW observables critically depend on the knowledge of the beam-spread at $Z^0$ pole, like the cross-section for $Z^0$ production, $Z^0$ total width and mass. We have found the following requirements on the beam spread at the CEPC $Z^0$ pole: 0.5%, 0.2% and 10%, respectively.

## 3. Method

As illustrated in Figure 2 [3], non-zero beam-spread will result in acolinearity of final state muons produced in $e^+e^- \rightarrow \mu^+\mu^-$. According to the expected performance of the central tracker at CEPC [1], muon polar angle resolution over the whole tracking volume should be 0.1 mrad, which corresponds to 100 μm position resolution in TPC. The effective center-of-mass energy $s'$ can be calculated from the reconstructed muons' polar angles [3]:

$$\frac{s'}{s} = \frac{\sin\theta^+ + \sin\theta^- - |\sin(\theta^+ + \theta^-)|}{\sin\theta^+ + \sin\theta^- + |\sin(\theta^+ + \theta^-)|},$$

where $s$ stands for the nominal CM energy and $\theta^+$ and $\theta^-$ are polar angles of outgoing $\mu^+$ and $\mu^-$.

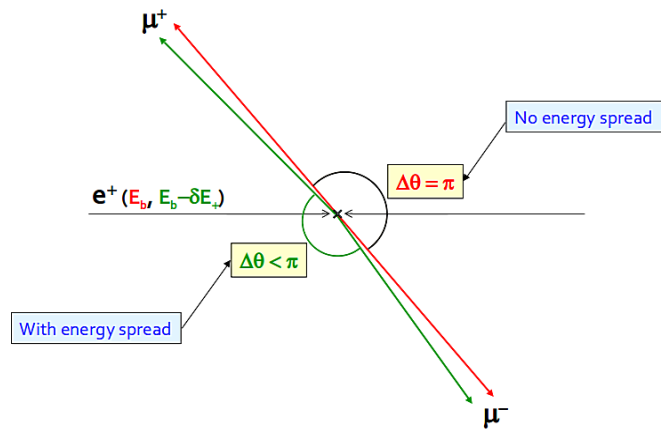

**Figure 2.** Process $e^+e^- \rightarrow \mu^+\mu^-$ without beam energy spread (red) and with beam energy spread (green).

In order to determine *s'* sensitivity to the beam-spread, we generated between 100K and 250K $e^+e^- \to \mu^+\mu^-$ events at 91.2 GeV and 240 GeV. Events are generated using WHIZARD 2.6.2 [4], in polar angle ranged from 8° to 172°, which is the angular acceptance of the TPC at CEPC [1]. Events are generated without any additional effects, in order to study individual effects of ISR, beamstrahlung and muon angular resolution competing with the beam-spread. Detector energy resolution is simulated by performing Gaussian smearing of the muons' polar angles in case of a few different central tracker reconstruction capabilities.

**4. Results**

Figure 3 illustrates *s'* distribution in the presence of ISR and the beam-spread. The beam-spread is assumed in accordance with the nominal CEPC beam parameters: 0.134% of the beam energy at 240 GeV and 0.08% of the beam energy at 91.2 GeV. Tracker (muon polar angle) resolution is here assumed to be infinitely accurate.

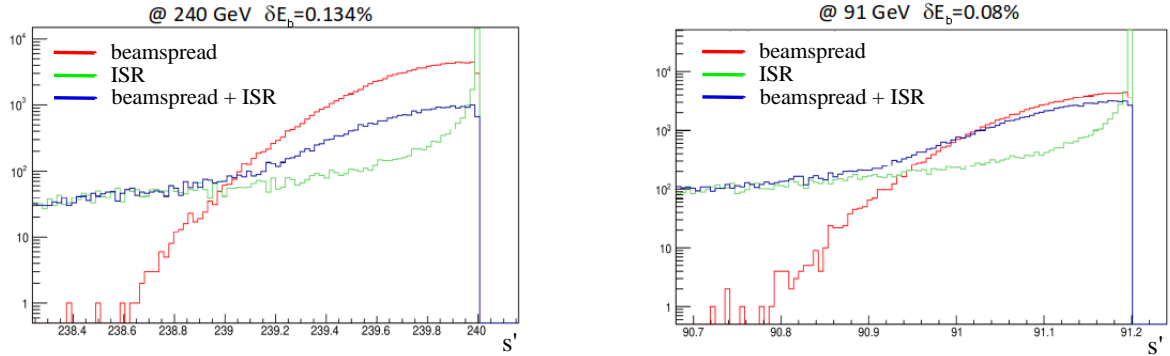

**Figure 3.** *s'* sensitivity to the beam spread at 240 GeV (left) and 91.2 GeV CEPC(right).

It can be seen that at both energies the beam-spread dominates the *s'* shape at energies close to the nominal center-of-mass energy. In order to rely on this method, excellent theoretical description of ISR effect is required. In Figure 4, the effect of muon polar angle resolutions of 0.1 mrad and 1 mrad are illustrated on top of ISR and the beam-spread.

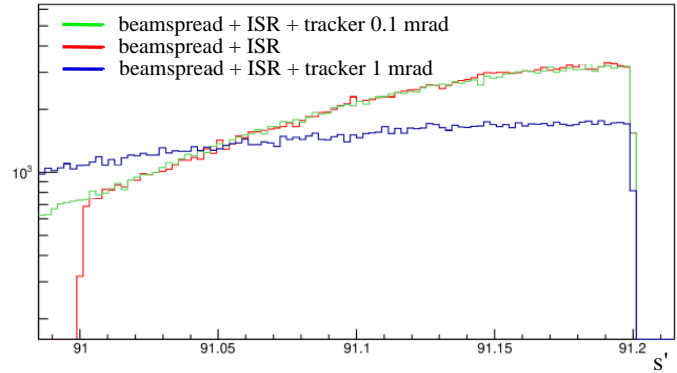

**Figure 4.** *s'* sensitivity to the beam spread and tracker resolution at 91.2 GeV.

From Figure 4 is clear that 0.1 mrad tracker resolution of muons polar angles reconstruction does not affect the *s'* sensitivity to the beam spread. On the other hand, tracker resolution of 1 mrad significantly influences the method. The same stands for 120 GeV beam.

Then it can be asked how far one may go in deterioration of the central tracker performance. The answer is shown at Figure 5.illustrating that central tracker polar angle (positioning) resolution should stay within the range of 0.1-0.2 mrad.

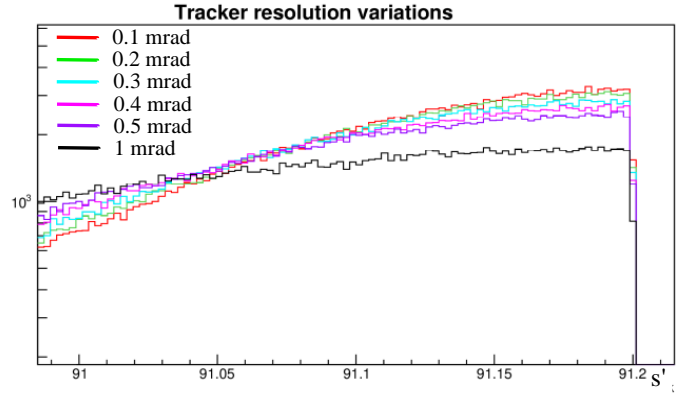

**Figure 5.** *s'* sensitivity to the different tracker resolutions in muon polar angle, simulated at 91.2 GeV.

Finally, the energy asymmetry of the colliding beams corresponding to the effective beam-spread can be demined from the population of the top-part of the *s'* distribution. Beam-spread values are varied around the nominal ones at 240 GeV and at the $Z^0$ pole. To reduce statistical uncertainties, 250K events is generated for each beam-spread value at 91.2 GeV and 100K events at 240 GeV. Results are shown at Figure 6.

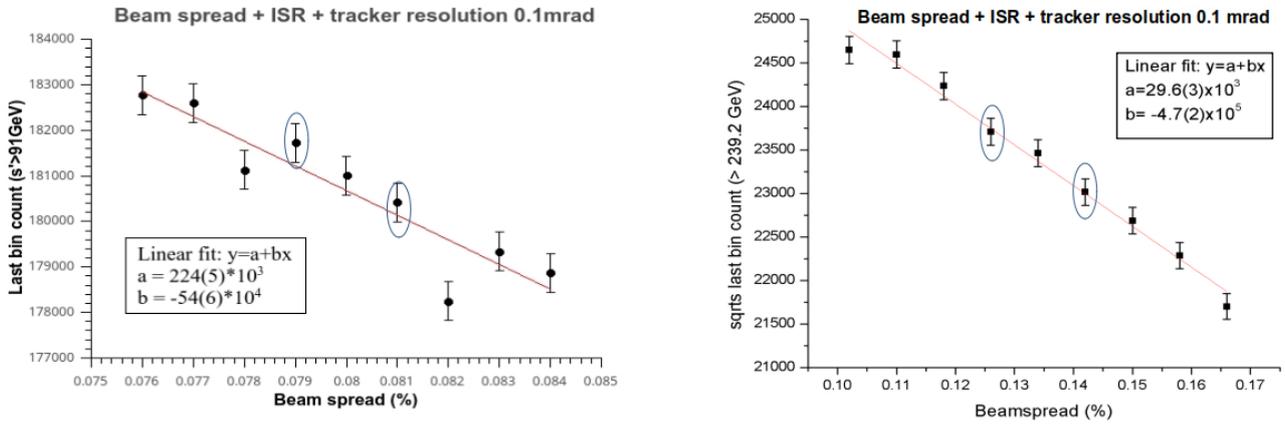

**Figure 6.** Dependence of the most energetic muons count on the beam-spread, at 91.2 GeV (left) and 240 GeV CEPC(right).

As expected, increase of the beam-spread leads to increase of acolinearity of outgoing muons, and the corresponding reduction of the center-of-mass energy available for a collision. The muon count dependence on the beam spread can be fitted using a simple linear fit. This fit enables us to calculate the effective beam-spread in the experiment simply by counting muons, as shown in Table 1. With a statistics of 250K events at $Z^0$ pole and 100K events at 240 GeV, relative variations of the nominal beam-spread of 2.5 % (15 %) can be measured, respectively.

For such relative precision in determination of the effective beam-spread, only 4 minutes of collecting the most energetic muons are needed at the $Z^0$ pole, with the CEPC nominal luminosity. Such a strict control of the beam-spread variation is not possible at 240 GeV CEPC. However, it is neither needed since for the luminosity uncertainty of $10^{-3}$, asymmetry in energy of the colliding beams should be known within 150% of the nominal beam-spread. The last row in Table 1 is given for comparison between CEPC and FCCee and shows the time needed to determine beam-spread variation at 91.2 GeV FCCee [3]. It reflects the combination of two compensating facts: instantaneous luminosity at FCCee is approximately an order of magnitude larger than at CEPC, while at CEPC $Z^0$ pole the beam-spread is almost two times smaller than at FCCee.

| CEPC | Luminosity @ IP (cm$^{-2}$ s$^{-1}$) | Nominal beam-spread (%) | Number of events | Cross-section $e^+e^- \to \mu^+\mu^-$ | Collecting time | Beam-spread ($\delta E_b$) variation |
|---|---|---|---|---|---|---|
| $Z^0$ pole | $3.2 \cdot 10^{35}$ | 0.080 | 250 KEvt. | 1.5 nb | ~ 4 min (2 min for $10^{-4}$ of $\Delta L/L$) | ~$2.5 \cdot 10^{-2} \cdot \delta E_b$ (900 keV) |
| Higgs factory | $3.0 \cdot 10^{34}$ | 0.134 | 100 KEvt. | 4.1 pb | ~ 10 days | ~ $0.15 \cdot \delta E_b$ (~24 MeV) |
| FCCee $Z^0$ pole | $2.3 \cdot 10^{36}$ | 0.132 | 540 KEvt. | 1.5 nb | ~ 3 min | ~$2 \cdot 10^{-3} \cdot \delta E_b$ (~120 keV) |

**Table 1.** Beam-spread variations experimentally accessible at CEPC and FCCee.

The luminosity precision requirement at the $Z^0$ pole versus the asymmetry of the colliding beam energies (effective beam-spread) can be met with only 2 min of running under CEPC nominal conditions.

## 5. Conclusion

The method of experimental determination of the beam-spread based on muon reconstruction from $e^+e^- \to \mu^+\mu^-$ nicely works at CEPC $Z^0$ pole, due to the high cross-section for di-muon production and high instantaneous luminosity. At $Z^0$ pole CEPC, 2.5% relative accuracy of the beam spread is feasible (i.e. < 1 MeV) after 4 minutes of data collection, while only 2 minutes of running are required to meet the relative precision of integrated luminosity uncertainty of $10^{-4}$. At 240 GeV, beam-energy asymmetry within the existing beam-spread is satisfactory for $10^{-3}$ precision goal on integrated luminosity.

Method requires further refinements to be applied: effect of ISR (theoretical) uncertainty, full detector simulation and impact of similar final states backgrounds and presence of beamstrahlung should be included in the future. Also, different choices of the fit function describing beam-spread dependence of the high-energy muons count, lead to the systematic uncertainty of the method, as well as the fact that the beam energy spread is not ideally Gaussian.